\documentclass[singlecolumn,article,superscriptaddress, longbibliography, nofootinbib]{revtex4-1}

\usepackage{hyperref}
\usepackage{url}

\usepackage{amsmath,amsfonts,amssymb}
\usepackage{graphicx}
\usepackage{color}
\usepackage{ulem}
\usepackage{float}

\begin{document}

\title{Selective Exposure shapes the Facebook News Diet}
\author{Matteo Cinelli}
\email[]{}
\affiliation{Applico Lab, CNR-ISC}

\author{Emanuele Brugnoli}
\email[]{}
\affiliation{Applico Lab, CNR-ISC}

\author{Ana Lucia Schmidt}
\email[]{}
\affiliation{Universit\'a di Venezia "Ca' Foscari"}

\author{Fabiana Zollo}
\email[]{}
\affiliation{Universit\'a di Venezia "Ca' Foscari"}
\affiliation{Applico Lab, CNR-ISC}

\author{Walter Quattrociocchi}
\email[]{}
\affiliation{Universit\'a di Venezia "Ca' Foscari"}
\affiliation{Applico Lab, CNR-ISC}

\author{Antonio Scala}
\email[]{}
\affiliation{Applico Lab, CNR-ISC}
\affiliation{LIMS, the London Institute for Mathematical Sciences}

\begin{abstract}

The social brain hypothesis fixes to 150 the number of social relationships we are able to maintain. Similar cognitive constraints emerge in several aspects of our daily life, from our mobility up to the way we communicate, and might even affect the way we consume information online. Indeed, despite the unprecedented amount of information we can access online, our attention span still remains limited.
Furthermore, recent studies showed the tendency of users to ignore dissenting information but to interact with information adhering to their point of view. 
In this paper, we quantitatively analyze users' attention economy in news consumption on social media by analyzing 14M users interacting with 583 news outlets (pages) on Facebook over a time span of 6 years.
In particular, we explore how users distribute their activity across news pages and topics. We find that, independently of their activity, users show the tendency to follow a very limited number of pages. On the other hand, users tend to interact with almost all the topics presented by their favored pages. 
Finally, we introduce a taxonomy accounting for users behavior to distinguish between patterns of selective exposure and interest.

Our findings suggest that segregation of users in echo chambers might be an emerging effect of users' activity on social media and that selective exposure -- i.e. the tendency of users to consume information interest coherent with their preferences -- could be a major driver in their consumption patterns.
\end{abstract}

\maketitle

\section{Introduction}

The social brain hypothesis fixes to 150 the number of social relationships we are able to maintain \cite{dunbar1998social,dunbar1992neocortex}. Such a theoretical cognitive limit emerges in several other contexts \cite{dunbar2012social} from the patterns of human mobility  \cite{alessandretti2018evidence} up to the way we communicate \cite{saramaki2014persistence,goncalves2011validation}.
Furthermore, the uptake of social media radically changed the way we consume content online. Indeed, the way we consume information and the cognitive limits and algorithmic mechanisms underpinning them bears on foundational issues concerning our news consumption patterns. As a consequence, in 2017 the World Economic Forum raised a warning on the potential of social media to distort the perception of reality\footnote{http://reports.weforum.org/global-risks-2017/part-2-social-and-political-challenges/2-1-western-democracy-in-crisis/}; possibly, such risk is related to the fact that social media induced a paradigm shift in the way we consume information~\cite{carlson2018facebook,quattrociocchi2014opinion}. Along this path, recent studies targeting Facebook ~\cite{zollo2017debunking,del2016spreading,bakshy2015exposure} showed that content consumption is dominated by selective exposure -- i.e.,  the tendency of users to ignore dissenting information and to interact with information adhering to their preferred narrative. 

Selective exposure may lead to the emergence of echo-chambers ~\cite{bastos2018geographic,del2016echo} -- i.e., groups of like-minded people cooperating to frame and reinforce a shared narrative -- thus facilitating fake news and more generally misinformation cascades~\cite{mocanu2015collective,bessi2015science}, especially since we switched from a paradigm where information was supplied by few official news sources mediated by experts and/or journalists, to the current disintermediated environment composed by a heterogeneous mass of information sources. Social media play a pivotal role not only in our social lives, but also in the political and civic world, coming to such an extent that they have rapidly become the main information source for many users~\cite{sehl2016public}. Essentially, online confirmation bias seems to account for users' decisions about consuming and spreading content; at the same time, the aggregation of favored information within those communities reinforces selective exposure and group polarization~\cite{sunstein2002law, quattrociocchi2017inside}.

Several works addressed the dynamics of news consumption through social media~\cite{allcott2017social, oeldorf2015posting, del2017mapping} and explored the interplay between selective exposure and political polarization on the Internet \cite{garrett2009politically, garrett2009echo}. Focusing on news consumption on social media, in~\cite{schmidt2017anatomy} the authors find that users' consumption patterns seem to determine the emergence of a sharp community structure among news outlets.
Nowadays, the understanding of the impact of social media on the news business model is one of the most pressing challenges for both science and society~\cite{flaxman2016filter,garimella2018quantifying,burgess2018youtube}.

In this paper, we perform a thorough quantitative analysis to characterize users' attention dynamics on news outlets on Facebook. 
In particular, we study how 14 million Facebook users distribute their activity among 50000 posts, clustered by topics, produced by 583 pages listed by the European Media Monitor on a 6 years time span.
We find that users, independently of their activity and of the time they spend online, show a tendency to interact with a very limited number of news outlets. 
To test the presence of selective exposure, whose evidence emerges from users focusing their attention on a set of preferred news sources, we analyze how homogeneously users distribute their activity across pages and topics. 
More precisely, the concentration of the distribution of likes towards a certain page or topic signals the presence of selective exposure, while the heterogeneity of such a distribution determines the strength of selective exposure.
We find that highly engaged users tend to concentrate their activity on few pages while being less selective to the topics presented by the pages. In general, we observe that selective exposure increases in strength when the activity of users (i.e. the number of likes) grows but is not affected by users' lifetime (i.e. the time span between the first and the last like).
Finally, we provide a taxonomy to classify users by means of their consumption patterns.
Our results suggest that the tendency of users to limit their attention on few news sources might be one of the factors behind the emergence of echo chambers online. 
The emerging outcome still underlines the tendency of users towards segregation, partly because of their attitude and cognitive limits, and partly because of the features of the social media in which they operate.

The paper is structured as follows. First, we describe the way users interact with posts, pages and topics, characterizing their news consumption habits. Then, we analyze users' attention patterns on pages and topics and discuss the mechanism of selective exposure as a quantitative heterogeneity problem. Finally, we conclude the paper by outlining a taxonomy of the users based on the comparison between their attention patterns with respect to pages and topics.

\section{Results and Discussion}
\subsection{Users' News Consumption}
\label{sec:Dstat}

News appear on Facebook as posts and users can interact with such posts through different actions, namely likes, comments and shares. A like is usually a positive feedback on a news item. A share indicates a desire to spread a news item to friends. A comment can have multiple features and meanings and can generate collective debate.
Since our aim is to investigate the mechanism of selective exposure we focus our analysis on the likes of the users, i.e on their positive feedback towards certain posts. As shown in previous works~\cite{schmidt2017anatomy}, likes are a good proxy of the users' activity in terms of engagement and attention patterns.

The interaction between users and posts can be represented as a bipartite network $G_{up}$, undirected and unweighted, in which the first partition has $n_u$ elements (corresponding to the users) while the second partition has $n_p$ elements (corresponding to the pages). The matrix $I_{up}$ representing such bipartite network is binary since a user is allowed to put one like per post; thus, we have $I_{up}=1$ if user $u$ likes post $p$, $0$ otherwise. Given $G_{up}$, the activity -- i.e. the number of likes -- of the user $u$ can be quantified by his/her degree $k_u = \sum_{p=1}^{n_p}I_{up}$.

In order to investigate the relationship between user and pages, from the bipartite network $G_{up}$ we obtain a second bipartite network with $n_u$ users and $n_{P}$ pages called $G_{uP}^*$, in which posts are simply grouped by the page that generated them. On such a network the activity of the user remains unchanged and the number of likes of user $u$ to page $P$ can be obtained as $I_{uP}^*=\sum_{p\in P}I_{up}$.

\begin{figure}[ht]
\begin{center}
\includegraphics[trim=0cm 2cm 0cm 2cm, clip=true, height=4 cm, width=6 cm]{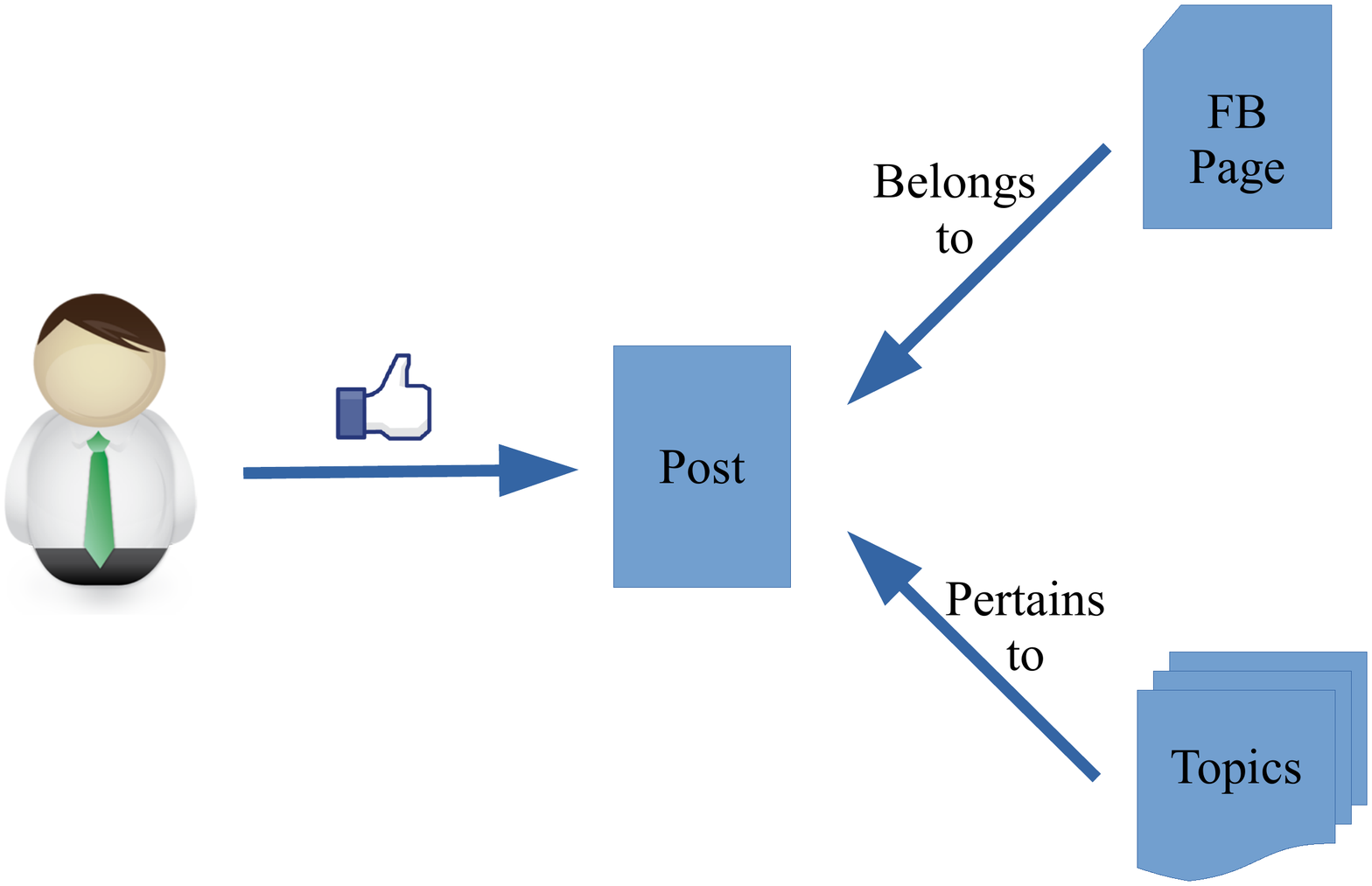}
\end{center}
\caption{Schematic representation of users' activity on Facebook for what concerns the analyses conducted in this paper. A user likes a post; each post belongs to a single Facebook page (top) and is related to one or more topics (bottom) that are attributed to the post via the topic modeling algorithm of \protect{~\cite{gerlach2018network}} . 
Single icons created using the free gallery of OpenOffice.}
\label{fig:pic_u}
\end{figure}

Additionally, the posts of the user-post network $G_{up}$ can be also grouped by the topic they treat using a topic modeling algorithm~\cite{gerlach2018network} as described in Section~\ref{sec:Methods}. Aggregating $G_{up}$ by topic, we generate a third bipartite network called $G^\dagger_{ut}$ with $n_u$ users and $n_t$ topics. A post can be considered a mixture of topics, all appearing in a certain proportion, and  the weighted bipartite network $G^\dagger_{ut}$ is represented by the matrix $I^\dagger_{ut}$ in which the weight of each element is proportional to the overall presence of a certain topic in the posts liked by a certain user (see in Section~\ref{sec:Methods}). Using $I^\dagger_{ut}$ we can study the activity of users with respect to different topics. A pictorial representation of the activity of the user and the relationship between posts, pages and topics in reported in Figure~\ref{fig:pic_u}.

\begin{figure}[ht!]

\includegraphics[trim=0cm 0cm 0cm 0cm, clip=true,scale = 0.25]{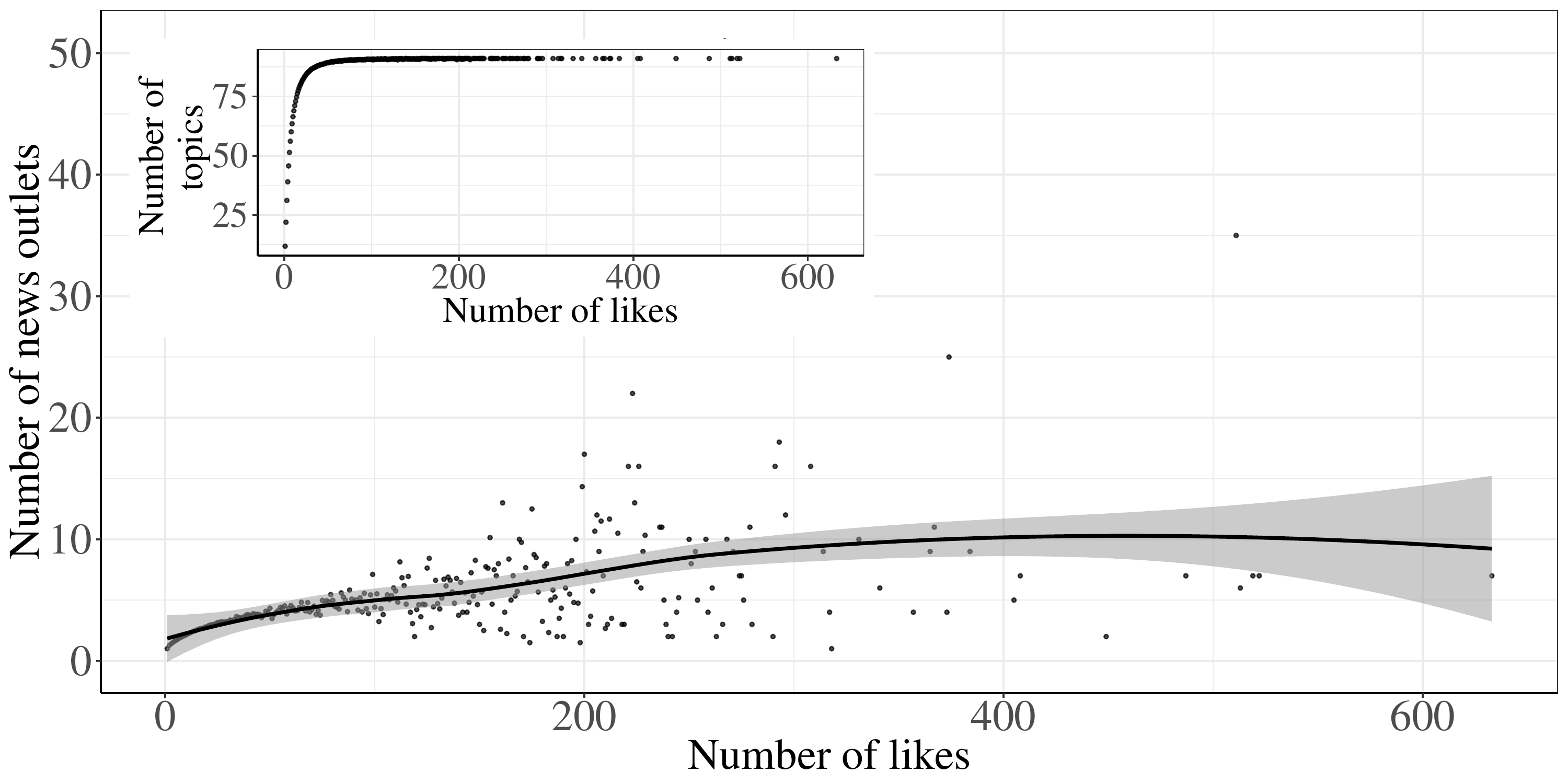}
\hfill
\includegraphics[trim=0cm 0cm 0cm 0cm, clip=true,scale = 0.25]{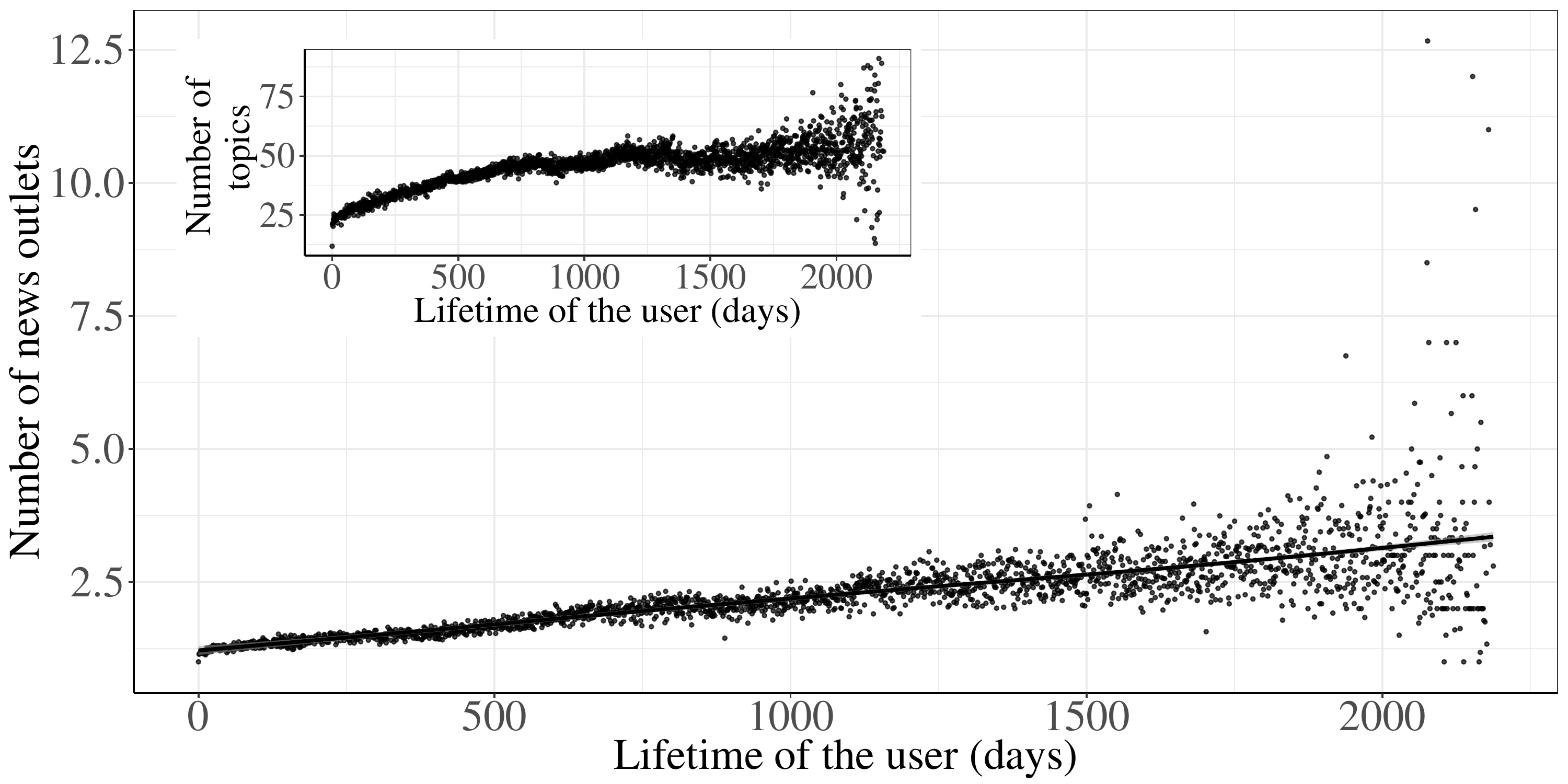}

\caption{Correlations between users' activity/lifetime and user engagement with pages and topics.
Left panel:  relationship between the average number of pages followed by users with respect to their activity (quantified by the number of likes). We observe that the average number of pages reaches a plateau of $\sim 10$ pages for users with an activity of more than $\sim 300$ likes. In the inset of the panel, we show the relationship between the average number of topics covered by users with respect to their activity. We observe that users with an activity of more than $\sim 10$ likes already reach a plateau  corresponding to the overall number of topics that is 91 (as explained in Section~\ref{sec:Methods}).
Right panel: relationship between the average number of pages followed by users with respect to their lifetime (quantified by the time between the first and the last like). We observe that the average number of pages grows slowly and reaches a value of $\sim 3$ pages for the most lifelong users. The inset of the right panel shows the relationship between the average number of topics and the users' lifetime. We observe that for lifetime larger than $\sim 1000$ days the number of topics reaches a value of $\sim 50$, corresponding roughly to $50\%$ of the overall topics. Further details are reported in Supplementary Information.  
}
\label{fig:Desc}
\end{figure}

Figure~\ref{fig:Desc} shows the average number of pages liked by users with respect to their activity and lifetime, whereas the former is defined as the number of likes of the user and the latter as the time span between the first and the last like a user put on two different posts.
In Figure~\ref{fig:Desc} (\textit{left}) we observe the relationship between the users' activity and the number of pages they interact with.
We notice that the average number of pages liked by a user reaches a plateau with increasing activity; in particular, users with more than $\sim 300$ likes concentrate, on average, their activity on only $\sim 10$ pages (for further details, see SI). 
This may be due to different -- an possibly co-interacting -- factors, such as the different narratives adopted by the pages in order to report information, the presence of natural limits to attention of the users, or even to the filtering due to the ranking algorithms used in the information search. 

To define the topics of the posts, we first pre-process the posts to extract the set of meaningful words $W$ (see sec.\ref{sec:Methods})  and then define the bipartite network $G^\diamond_{pw}$ that links each post $p$ to the words $w$ used in the post. 
We then apply the hierarchical stochastic block-modeling algorithm of ~\cite{gerlach2018network} (a well assessed topic modeling algorithm that takes a bipartite network as input) on $G^\diamond_{pw}$ to detect the topics  and find $91$ different topics (see section~\ref{sec:Methods}). We observe  that, since the analyzed pages are news outlets, most pages tend to cover almost all the topics (see SI).

The inset in Figure~\ref{fig:Desc} (\textit{left}) shows the number of topics an user interacts with respect to his/her activity.  At difference from what observed in the interaction with pages, users tend to interact with many topics regardless of their activity. In particular, users with more than $\sim 10$ likes already tend to interact with almost all the topics. Such a interaction patterns could be explained assuming that users tend to interact with all the topics presented by their reference pages. 

In Figure~\ref{fig:Desc} (\textit{right}) we notice that average number of pages users interact with grows slowly with the users' lifetime. However, the average number of topics reaches a plateau corresponding to $\sim 50\%$ of the overall topics for users with a lifetime larger than $\sim 1000$ days.

\subsection{Attention Patterns on Topics}
\label{subsec:TopicSelection}

Selective exposure relates to the tendency of users to concentrate their activity on specific topics or pages while ignoring other ones. For instance, a user who focuses his/her activity on a single topic (or page) would display higher selective exposure than a user who interacts with multiple topics.
Focusing on a single topic rather than on different ones entails an heterogeneity in the distribution of the user's activity that can be directly associated with the mechanism of selective exposure.

Therefore, a good proxy to selective exposure is a measure able to quantify heterogeneity in the distribution of users' activity across different elements, namely topics or pages.

The Gini index~\cite{gini1921measurement} is a classic example of synthetic indicator for measuring inequality of social and economic conditions~\cite{xu2003has}; hence, to give a measure of selective exposure, we apply the Gini index on the users' activity on different topics as stored on the rows of the weighted incidence matrix $I^\dagger_{ut}$. Notice that, consistently with the use of a state of the art topic modeling algorithm~\cite{gerlach2018network}, a post is considered a mixture of topics all appearing in different proportions. Consequently, the interaction of a user with multiple topics, that derives from liking one or more posts treating that topics, is still consistent with a mixed membership model~\cite{airoldi2014handbook}.

The Gini index can be defined starting from the Gini absolute mean difference $\Delta$~\cite{kendall1958theadvanced} of a generic vector $y$ with $n$ elements that can be written as:
\begin{equation}
\Delta = \frac{1}{n^2}\sum_{i=1}^n \sum_{j=1}^n |y_{i} - y_{j}|
\label{eq:delta}
\end{equation}
The relative mean difference is consequently defined as $\Delta/\mu_y$ where $\mu_y= n^{-1}\sum_{i=1}^n y_i$. Thus, the relative mean difference equals the absolute mean difference divided by the mean of the vector $y$.  The Gini index $g$ is one-half of the Gini relative mean difference~\cite{anand1983inequality}
\begin{equation}
g = \frac{\Delta}{2 \mu_y}
\label{eq:Gini}
\end{equation}

We estimate the strength of selective exposure of user $u$ to topics using, accordingly to Equation~\ref{eq:Gini}, using the following expression of the Gini index:
\begin{equation}
    g^\dagger = \frac{1}{2 \, n_t} \frac{\sum_{t=1}^{n_t}\sum_{q=1}^{n_t} |I_{ut}^\dagger -I_{uq}^\dagger|}{\sum_{t=1}^{n_t} I_{ut}^\dagger}
    \label{eq:Gini_t}
\end{equation}

\begin{figure}
\centering
\begin{minipage}{0.45\textwidth}
\centering
\includegraphics[trim=0cm 0cm 0cm 0cm, clip=true,scale = 0.28]{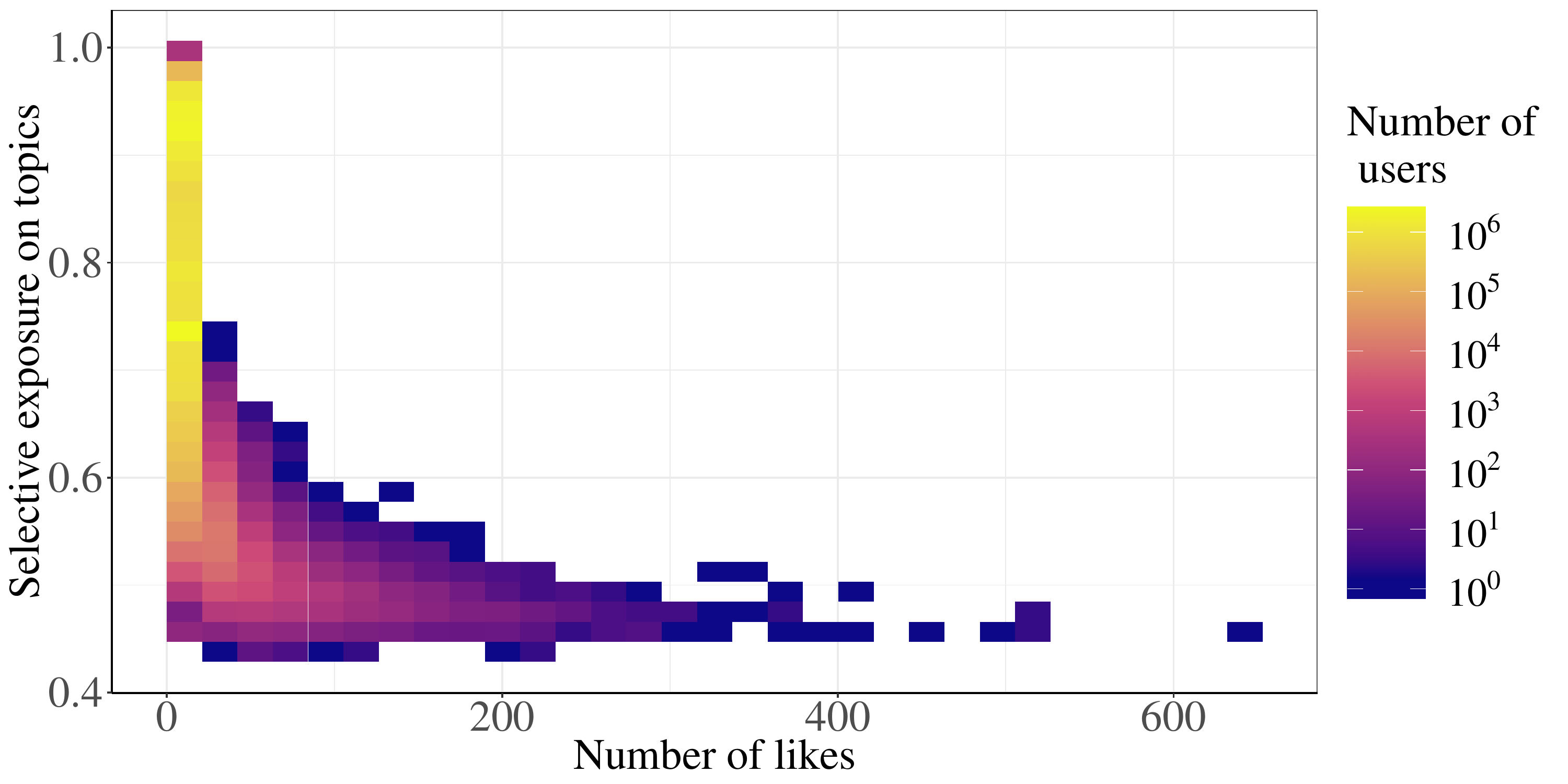}
\end{minipage}
\begin{minipage}{0.45\textwidth}
\centering
\includegraphics[trim=0cm 0cm 0cm 0cm, clip=true,scale = 0.28]{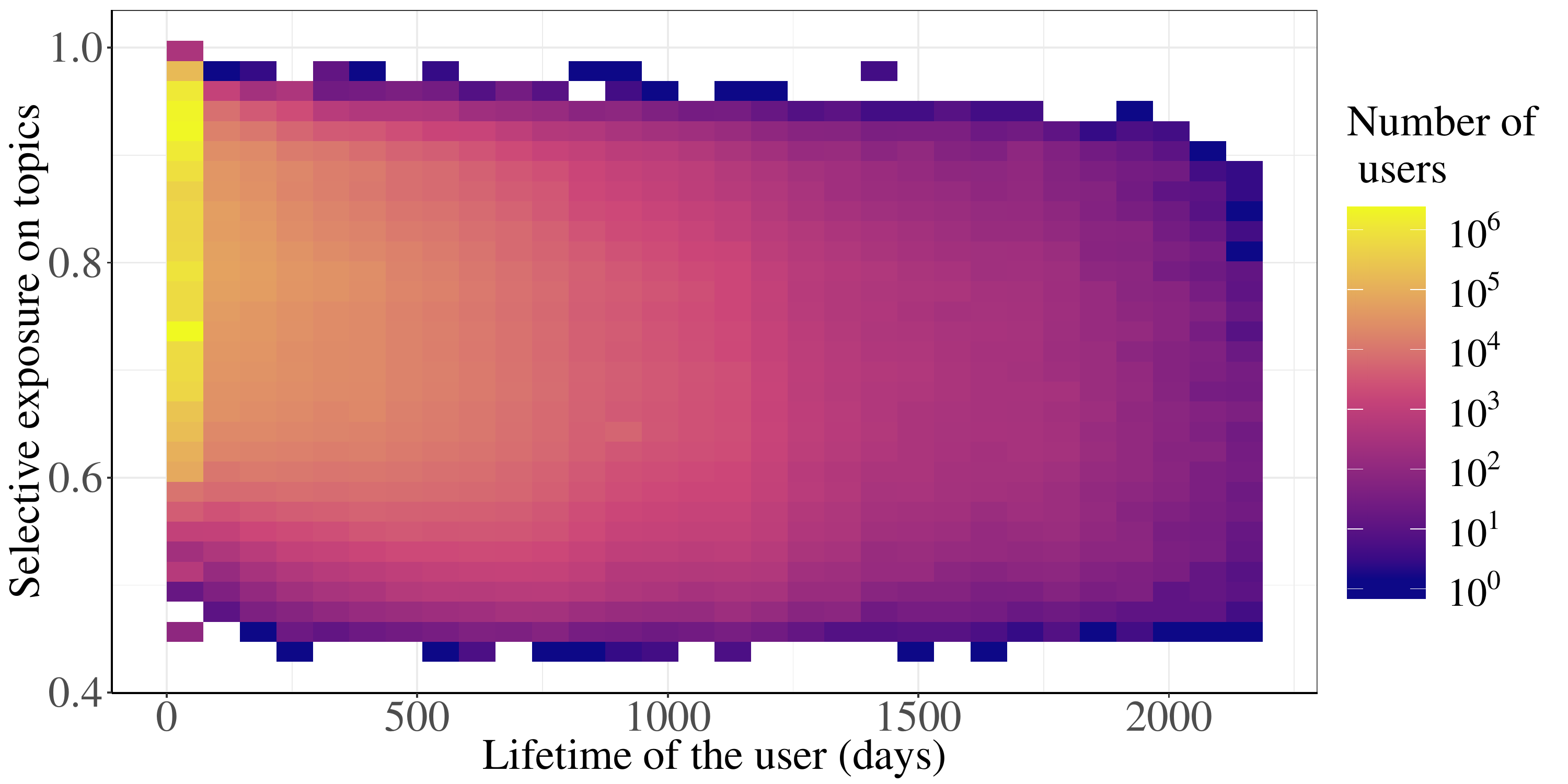}
\end{minipage}
\caption{Correlation among users’ selective exposure (as measured by the Gini coefficient $g^\dagger$) and their activity/lifetime.  The color scale of the distributions represents the amount of users related to a certain $(x, y)$ couple.  Left panel:  the distribution of  selective  exposure  to  topics with  respect  to  users’  activity  shows  that  to  increasing  activity  levels  correspond  lower  selective exposure, i.e.  users concentrate on a higher number of topics.  Right panel:  the distribution of selective exposure to topics with respect to users’ lifetime (measured in days) shows that the mechanism of choice of topics does not seem to be influenced by the time the users have been present on the social medium. Further details are reported in Supplementary Information.}
\label{fig:selex_t}
\end{figure}
Values of $g^\dagger_u \sim 1$ signal that the user $u$ concentrates his/her activity on few topics, while values of $g^\dagger_u \sim 0$ signal the tendency to be active on different topics.
In Figure~\ref{fig:selex_t}, we show the strength of selective exposure (as measured by the Gini index $g^\dagger_u$) with respect to the users' activity and lifetime.  We observe that to increasing values of activity corresponds a progressively weaker selective exposure; on the other hand, users' lifetime does not show strong correlations with their focus on specific topics. This result is consistent with the fact that Facebook pages tend to span several topics (see SI) and that highly active users are more likely to consume a wider amount of topics, thus decreasing their selective exposure to topics. 
In fact, even if users never get to a flat "diet" on topics (corresponding to a Gini index $\sim 0$), we note that users consume more topics with increasing activity, i.e. most active users are those with the weaker selective exposure to topics.

\subsection{Attention Patterns on Pages}
\label{subsec:PageSelection}

To understand whether the mechanism of selective exposure to pages -- if present -- could be different from that observed for topics, we replicate the analysis of   Section~\ref{subsec:TopicSelection} by considering the matrix $I^*_{uP}$, i.e. considering the interaction of users with news outlets (pages).

In this case, the expression for the Gini index $g_u$ of the user $u$ with respect to pages he/she likes is:
\begin{equation}
    g^* = \frac{1}{2\, n_P}\frac{\sum_{P=1}^{n_P}\sum_{Q=1}^{n_P} |I^*_{uP}-I^*_{uQ}|}{\sum_{P=1}^{n_P} I^*_{uP}}
    \label{eq:Gini_p}
\end{equation}

However, applying the Gini index to our dataset would introduce a bias due to the sparsity of the matrix $I^*_{uP}$. In fact, we have many users whose activity is smaller than the number of pages (i.e. the sum of the entries of a row $u$ of $I^*$ is often much smaller than than the number of columns  $n_P$). In such cases, the Gini index displays a bias towards high values~\cite{bernasco2017more} of $g^*$ (see SI) since the denominator of Equation \ref{eq:Gini_p} is small and the possibility of perfect equidistribution -- i.e. the same number of likes on each page -- cannot be achieved. Therefore, to avoid such flaw of the Gini index in the case of sparse data, we renormalize the Gini index according to the minimum and maximum values it can assume :
\begin{equation}
    g^{\triangleright} = \frac{g^* - g^*_{min}}{ g^*_{max}- g^*_{min}}
    \label{eq:Gini_norm}
\end{equation}
where $g^*_{max}=1$ is the maximum value of the Gini coefficient, while $g^*_{min}$ is the minimum value of the Gini coefficient. As shown in Section \ref{subsec:minGini}, $g^*_{min}$ depends on the number of likes $n_l$ and the number of pages $n_P$ ;  when $n_l<n_P$, due to the "not enough data bias" we have that $g^*_{min}>0$. 
Thus, the quantification of selective exposure can be assessed using the normalized Gini index $g^{\triangleright}$ as in Equation~\ref{eq:Gini_norm}.

\begin{figure}[h!]
\includegraphics[trim=0cm 0cm 0cm 0cm, clip=true,scale = 0.28]{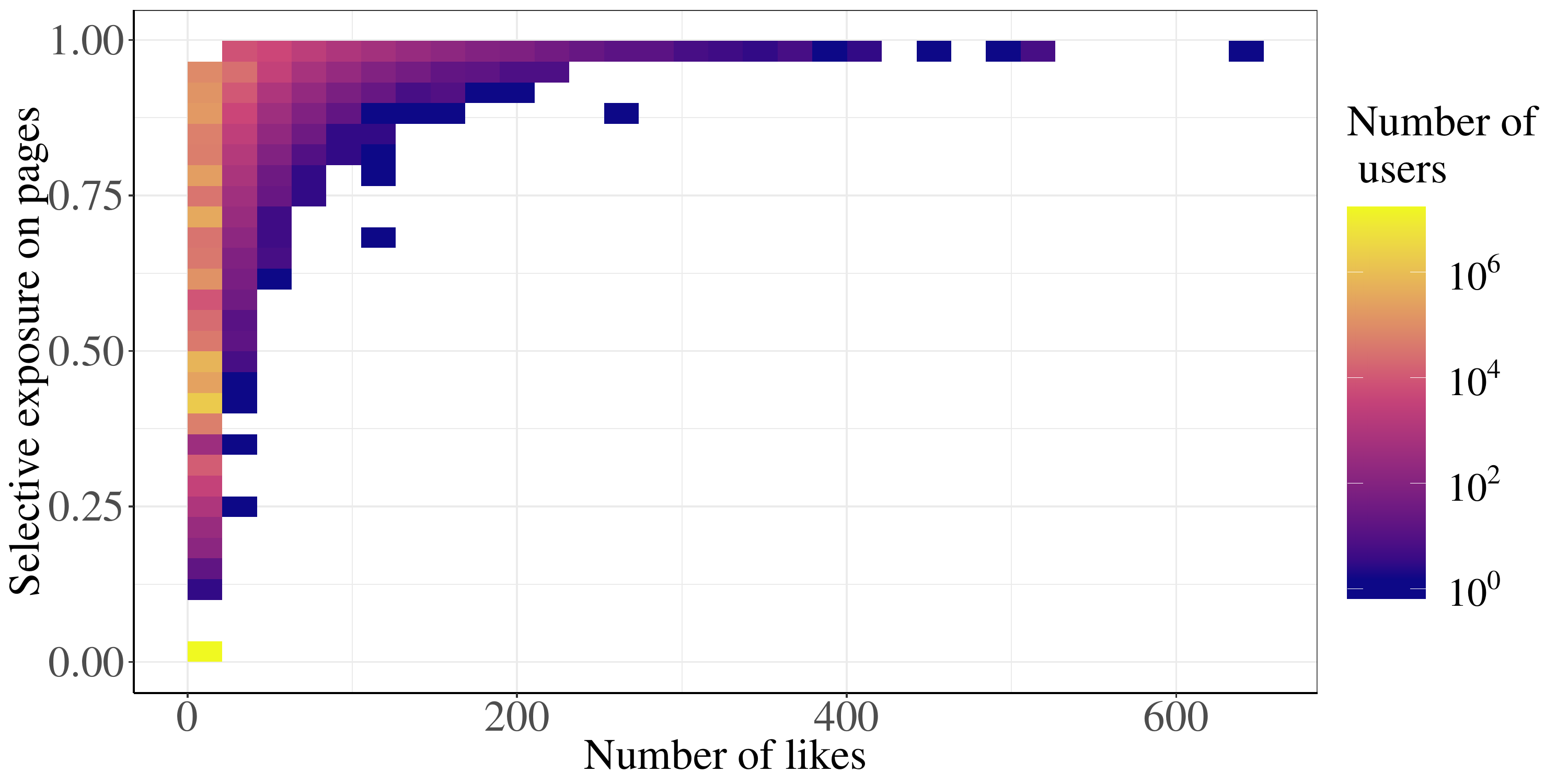}
\hfill
\includegraphics[trim=0cm 0cm 0cm 0cm, clip=true,scale = 0.28]{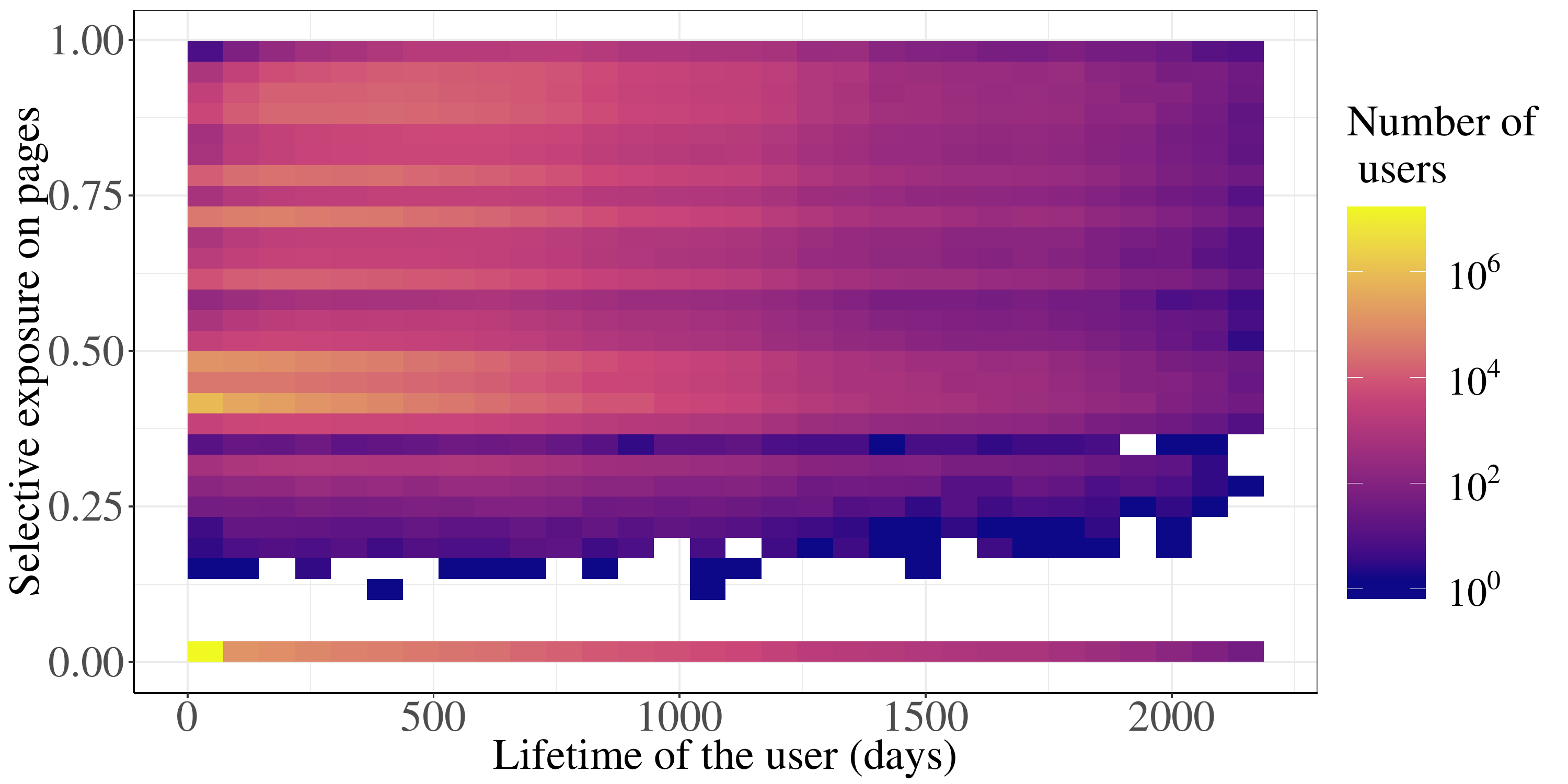}
\caption{Correlation among users' selective exposure (as measured by the Gini coefficient $g^\triangleright$) and their activity/lifetime. The color scale of the distributions represents the amount of users related to a certain $(x,y)$ couple.
Left panel: the distribution of selective exposure to pages with respect to users' activity shows that to increasing activity levels correspond higher selective exposure, i.e. users concentrate on fewer pages.
Right panel: the distribution of selective exposure to pages with respect to users' lifetime (measured in days) shows that the mechanism of choice of pages does not seem to be influenced by the time users have been present on the social medium.  The results of the panels are consistent with a way of choosing news outlets based on selective exposure rather than on confrontation among several sources; it is also consistent with a reinforcement mechanism for which the higher the activity, the stronger the concentration on fewer pages.
Further details are reported in Supplementary Information.}
\label{fig:selex_p}
\end{figure}

In Figure~\ref{fig:selex_p} we observe that the mechanism of selective exposure is present also in the case of pages, but with a completely different trend than what observed in the case of topics. 
In fact, we observe that to increasing values of activity correspond a concentration of users toward high values of  $g^{\triangleright}_u$, i.e. users' selective exposure to pages increases. On the other hand, users' lifetimes do not show strong correlations with $g^{\triangleright}_u$; hence, the mechanism of choice of pages does not seem to be influenced by the time users has been present on the medium.  Such results are consistent with a way of choosing news outlets based on selective exposure rather than on confrontation among several sources; it is also consistent with a reinforcement mechanism for which the higher the activity, the stronger the concentration on fewer pages.
In other words, we observe that users, especially the most active, tend to affiliate to pages and to their narratives regardless of the topics they treat. What appears is that the consumption of news depends on very few sources of information and could be almost independent on the subjects treated.

\subsection{Comparing Activity on Pages and Topics}

In this section we compare the two mechanisms of selective exposure. Indeed, users can display different profiles of selective exposure with respect to pages and topics and the knowledge of both dimensions can be helpful in order to characterize their attention patterns on social media.

\begin{figure}[h!]
\begin{center}
\includegraphics[trim=0cm 0cm 0cm 0cm, clip=true, scale = 0.4]{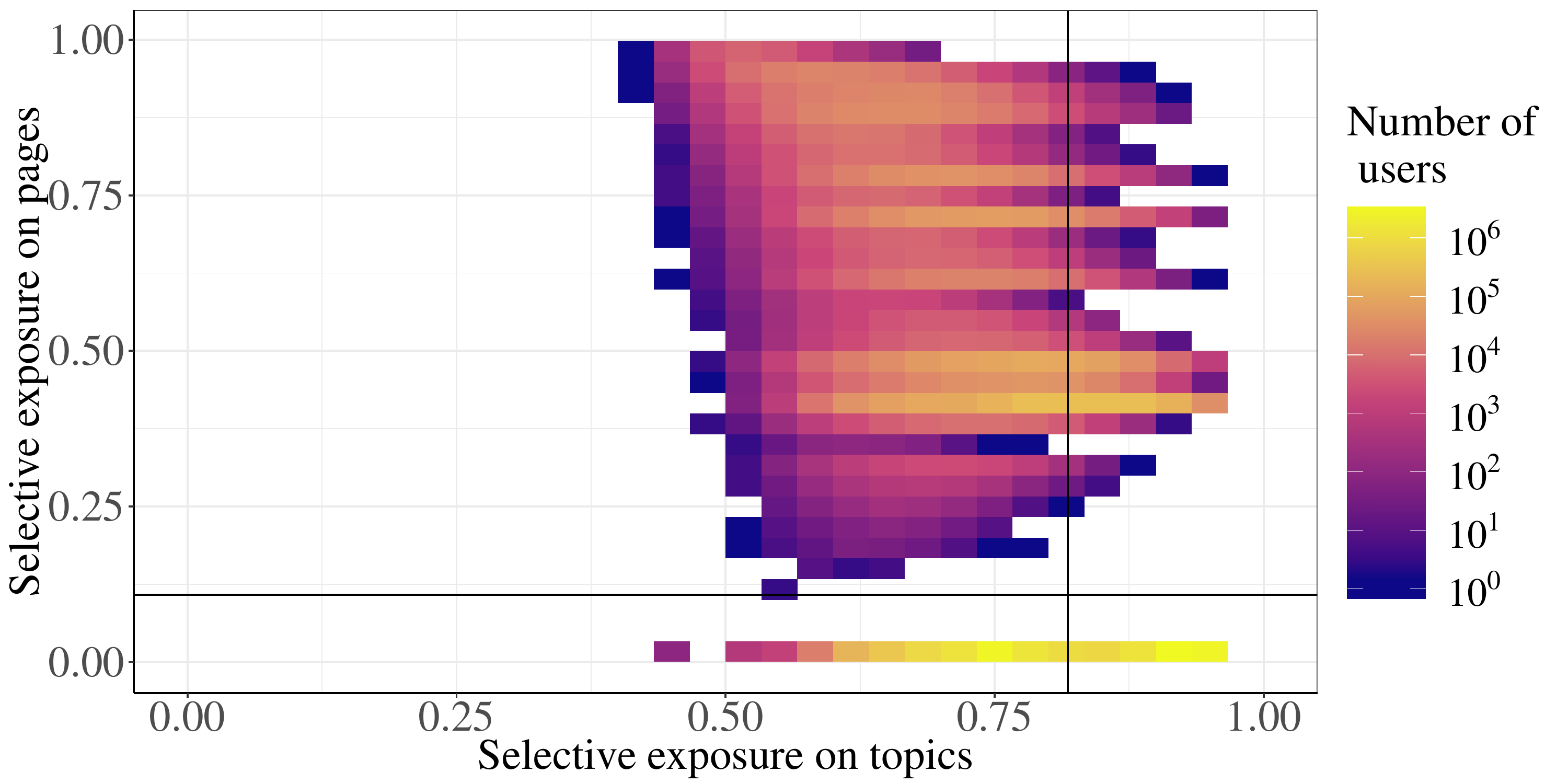}
\caption{Interrelation between the mechanism of selective exposure to pages and topics. The area is divided in four regions and three of them are labeled since they can be associated with different kind of selective exposure displayed by users. The regions determined by the average values of selective exposure to topics $g^\dagger = 0.818$ and topics $g^\triangleright = 0.108$. The region of multi-topic selective exposure is located top-left, the region of single-topic selective exposure is located top-right while the region of exposure by interest is located bottom left.
The color scale of the distribution represents the amount of users related to a certain $(x,y)$ couple.}
\label{fig:selex_comp}
\end{center}
\end{figure}

In Figure~\ref{fig:selex_comp}, by combining the results related to users' selective exposure to both pages and topics, we report different classes of users based on their statistical signatures. Users can be classified in three classes that are related to a specific type of selective exposure:

\begin{itemize}
    \item \textit{Multi-topic selective exposure}: high selective exposure to pages and low selective exposure to topics. Users in the region of multi-topic selective exposure are affiliated to one or few pages while spanning many topics. 
    \item \textit{Single topic selective exposure}: high selective exposure to pages and high selective exposure to topics. Users in the region of single-topic selective exposure are affiliated to one or few pages but they tend to focus their attention on specific contents.
    \item \textit{Exposure by Interest}: low selective exposure to pages and high selective exposure to topics. Users in the region of interest are not affiliated to pages but browse different sources consuming the contents they are interested in.
\end{itemize}

Figure~\ref{fig:selex_comp} is divided in four regions determined by the average values of selective exposure to pages and topics. Notice that the region of high selective exposure to topics / low selective exposure of pages corresponds mostly to users with very few likes; for such users the statistics is too low for assessing their interaction mechanisms with the social medium.

The region of multi-topic selective exposure is located top-left, the region of single-topic selective exposure is located top-right while the region of exposue by interest is located bottom right.
In Figure~\ref{fig:selex_comp} we observe that the largest fraction of users are located in the region of multi-topic selective exposure, accordingly with the fact that users tend to display a high selective exposure to pages and a low selective exposure to topics. The users with highest selective exposure to pages are also those with the highest activity (see Figure~\ref{fig:selex_p} and SI). Other users are located in the region of single-topic selective exposure meaning that they tend to focus on few pages and topics. We note that such users, having a high selective exposure to topics, also display an average activity that is lower than that of users in the region of multi-topic selective exposure (see Figure~\ref{fig:selex_t} and SI). 

The region of exposure by interest is well populated, however in such a region (as well as the fourth region located bottom left) the characterization of the behavior has to be carefully considered, since users users with low selective exposure to pages are also those with the lowest activity (see Figure~\ref{fig:selex_t},~\ref{fig:selex_p} and SI).

\section{Conclusions}\label{sec:Conclusion}
In this paper we explored the news diet of users on social media.
The economy of attention on social media is characterized by different features, one of which is selective exposure. Analyzing the interaction between 14 millions users and 583 news outlets, we find that users tend to interact with a very limited maximum amount of pages and that, similarly to a Dunbar number, it does not depend on their activity or lifetime. We find different features in the mechanism of selective exposure to pages respect to the mechanism of selective exposure to topics.
In particular, the probability of finding users with high selective exposure to pages increases with the users' activity, while in the case of topics, selective exposure decreases with activity. However, in both cases the lifetime of the user has no particular influence on the mechanism of selective exposure. 

By confronting the mechanisms of selective exposure to pages and topics, it is possible to differentiate between users' attention patterns to understand whether they are driven by selective exposure or interest.
Our findings suggest that the mechanism of selective exposure, together with users' limits to attention, strongly affects the way users select and consume news, and is likely to play a role in the segregation process that leads to the formation of echo chambers.
Further studies and datasets would be needed to investigate whether is the presentation priority of the news due to the Facebook algorithm is significantly relevant for the choice of news sources selected.

\section{Materials and Methods}\label{sec:Methods}

\subsection{Topic Modeling Algorithm}

Topic modeling consists in the application of machine learning tools to infer the latent topical structure of a collection of documents. 

Well-established and widely used topic models are probabilistic models, such as probabilistic Latent Semantic Analysis (pLSA)~\cite{hofmann1999probabilistic} and Latent Dirichlet Allocation (LDA)~\cite{blei2003latent}, an improvement of pLSA that exploits Bayesian statistics), where each document is a mixture of topics while each topic is a mixture of words.
Despite being the state of the art method for topic modeling, LDA suffers of several restrictions such as the risk of overfitting and the aprioristic choice of the number of topics~\cite{gerlach2018network}, among others~\cite{griffiths2005integrating,li2006pachinko,zhou2015negative}.
Such shortcomings of LDA have been recently addressed~\cite{gerlach2018network} by exploiting the conceptual relationship between topic modeling and community detection in networks.

By representing the relationship between words and documents (posts in our case) as a bipartite network, the algorithm proposed by~\cite{gerlach2018network} detects communities (i.e. cluster of densely interconnected nodes) using a hierarchical Stochastic Block Modeling (hSBM) algorithm~\cite{peixoto2014hierarchical, peixoto2015model, peixoto2017nonparametric}. The hSBM is a hierarchical version of the stochastic block model (SBM), a generative method for networks with block structure (i.e. communities) that serves as a base for community detection using statistical inference~\cite{holland1983stochastic, moore2017computer}.

In~\cite{gerlach2018network} a comparison between topic modeling and community detection algorithms, namely pLSA and SBM and LDA and hSBM, is carried on in order to demonstrate the suitability of hSBM for topic modeling problems. In particular, a mixed membership version of the SBM is formally proven to be equivalent to pLSA while the hSBM is shown to be conceptually similar to LDA.
In fact, hSBM represents a non parametric bayesian improvement of the SBM in the same way LDA is an improvement of pLSA based on bayesian statistics.

\subsection{Data Processing and Topic Modeling}

In our paper we exploit the hSBM algorithm for topic modeling on a bipartite network in which one partition is made up of 50000 pre-processed Facebook posts while the other is made up of the words contained in such posts. The raw Facebook posts that we consider are produced by a set of 583 news outlets. Such posts often include a link to an external website containing an article whose .html file is downloaded, parsed and reported as part of the raw post.
The raw posts are then processed in the following way: punctuation and stopwords are removed, words are lemmatised, part of speech tagging is executed keeping only nouns, posts with less than 5 words are removed.

\begin{figure}[ht]
\begin{center}
\includegraphics[trim=0cm 0cm 0cm 0cm, clip=true, height=6.5cm, width=8.5cm]{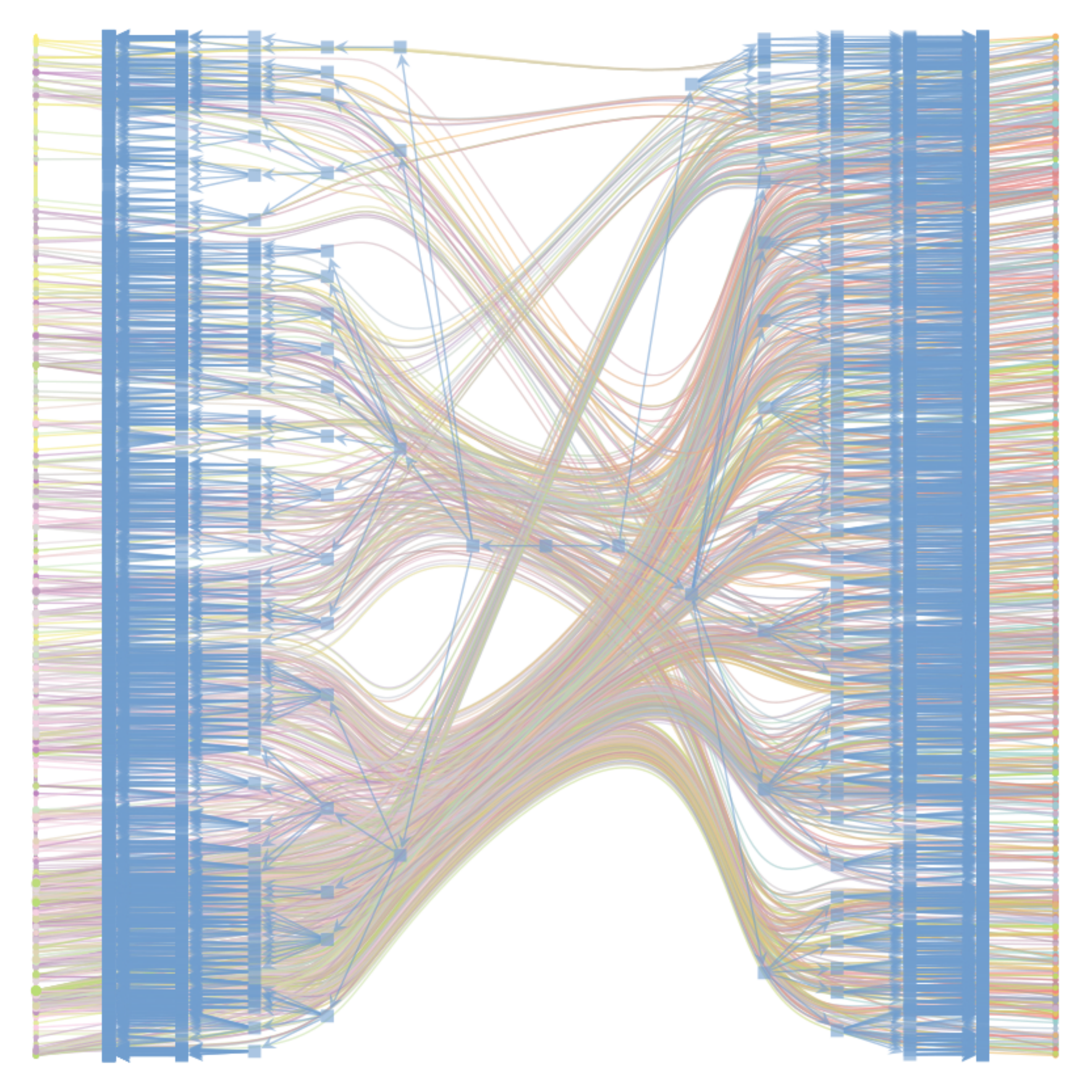}
\caption{Bipartite representation of the posts-words network. The partition on the left is made up of Facebook posts while the partition on the right is made up of words contained in such posts after pre-processing. The hierarchy obtained from hSBM is reported in blue and has 5 levels.}
\label{fig:river_t}
\end{center}
\end{figure}

After processing the text we run the hSBM algorithm on the considered network obtaining a hierarchy of topics with 5 levels, as displayed in Figure~\ref{fig:river_t}.
From the hierarchical structure of the topics we select the third level, containing 91 topics, for interpretability reasons. According to the results of the topic modeling algorithm each topic $t$ is present in each post $i$ in a certain proportion $p_t^i \in [0,1]$ and $\sum_{t=1}^{n_{t}} p_t^i = 1 \quad \forall i$.

In order to count the amount of topics related to each post (as displayed in the inset of Figure~\ref{fig:Desc}) we binarize the outcome of the topic modeling as follows. We assume that a post treats a topic if that post is associated to the topic more than the average association across all the considered posts, i.e. $ p^i > \frac{1}{n_{p}}\sum_{i=1}^{n_{p}}p_t^i$.
Consequently a topic $t$ is considered in the pool of topics liked by the user $i$ if he/she liked at least one post that contains the topic $t$.

\subsection{Minimum Gini coefficient}\label{subsec:minGini}

For simplicity, let's calculate the minimum value a Gini coefficient can attain by considering an user that puts $n_{likes}$ likes on $n_P$ pages. In this case, our Gini coefficient $g^*$ can be written as 
\begin{equation}
    g^* = \frac{1}{2\, n_P}\frac{\sum_{P=1}^{n_P}\sum_{Q=1}^{n_P} |I^*_{uP}-I^*_{uQ}|}{\sum_{P=1}^{n_P} I^*_{uP}}
\end{equation}

If the user has an overall activity greater than the number of pages, the coefficient $g_{min}^*\sim 0$ since a homogeneous distribution of likes $I_{uP}\sim n_{likes}/n_P$ across pages is allowed. On the other hand, when the overall number of likes is smaller than the number of pages, then the minimum value of the Gini index is in general greater than 0, since the likesa are concentrated only on $n_{likes}<n_P$ pages, i.e.  the distribution of likes is heterogeneous. Again, in this case  we can compute the lower bound $g^*_{min}$ to the Gini index, by supposing that the user spreads uniformly his/her likes over $n_{likes}$ pages (by putting 1 like per page); by substituting in eq.\ref{eq:Gini_p}, we obtain 
\begin{equation}
g^*_{min}=\frac{n_P - \sum_P I_{uP}}{n_P}
\end{equation}
, where $\sum_P I_{uP}{n_P}$ is the number of likes of the user. Summarising, the coefficient $g^*_{min}$ can be written as:
\begin{equation}
g^*_{min}=
    \begin{cases}
     \frac{n_P - n_{likes}}{n_P} & \text{if } n_{likes} \leq n_{P} \\
     0 & \text{otherwise } \\
    \end{cases}
\end{equation}

\subsection{Data Collection}

The Europe Media Monitor provides a list of all news sources. We limit our collection to Facebook pages associated to such sources reporting in English. The downloaded data from each page include all of the posts made from January 1, 2010 to December 31, 2015, as well as all of the likes and comments on those posts. In this paper we consider a sample of the original dataset made up of 50000 posts produced by 583 pages spanning the 6 years time window.



%

\end{document}